\documentclass[aps,amsfonts,amsmath,prd,preprint,nofootinbib]{revtex4}
\usepackage{graphicx}

\newcommand{\beq}{\begin{equation}}
\newcommand{\eeq}{\end{equation}}

\def\lap{\lower.5ex\hbox{$\; \buildrel < \over \sim \;$}}
\def\gap{\lower.5ex\hbox{$\; \buildrel > \over \sim \;$}}

\begin{document}

\title{Arrows of time and the beginning of the universe}

\author{Alexander Vilenkin}

\address{
Institute of Cosmology, Department of Physics and Astronomy,\\ 
Tufts University, Medford, MA 02155, USA}

\begin{abstract}

I examine two cosmological scenarios in which the thermodynamic arrow of time points in opposite directions in the asymptotic past and future.  The first scenario, suggested by Aguirre and Gratton, assumes that the two asymptotic regions are separated by a de Sitter-like bounce, with low-entropy boundary conditions imposed at the bounce.  Such boundary conditions naturally arise from quantum cosmology with Hartle-Hawking wave function of the universe.  The bounce hypersurface breaks de Sitter invariance and represents the beginning of the universe in this model.  The second scenario, proposed by Carroll and Chen, assumes some generic initial conditions on an infinite spacelike Cauchy surface.  They argue that the resulting spacetime will be non-singular, apart from black holes that could be formed as the initial data is evolved, and will exhibit eternal inflation in both time directions.  Here I show, assuming the null convergence condition, that the Cauchy surface in a non-singular (apart from black holes) universe with two asymptotically inflating regions must necessarily be compact.  I also argue that the size of the universe at the bounce between the two asymptotic regions cannot much exceed the de Sitter horizon.  The spacetime structure is then very similar to that in the Aguirre-Gratton scenario and does require special boundary conditions at the bounce.  If cosmological singularities are allowed, then an infinite Cauchy surface with `random' initial data will generally produce inflating regions in both time directions.  These regions, however, will be surrounded by singularities and will have singularities in their past or future.

\end{abstract}

\maketitle

\section{Introduction}

It has been recently shown \cite{Borde} that the spacetime of an inflationary universe is necessarily past-incomplete, even though inflation may be eternal to the future.  All past-directed timelike and null geodesics, except maybe a set of measure zero, reach the boundary of the inflating region of spacetime in a finite proper time (finite affine length, in the null case).  Unlike earlier singularity theorems \cite{HawkingEllis,WaldBook}, the theorem of Ref.~\cite{Borde} does not rely on Einstein's equations and does not assume any energy conditions.  To show the incompleteness of a given geodesic, all it requires is that the expansion rate averaged along the geodesic is greater than zero,\footnote{The expansion rate $H$ is defined in terms of a comoving congruence, and the averaging is performed over proper time (affine length) along the geodesic; see \cite{Borde} for details.}
\beq
H_{av} > 0.
\label{H>0}
\eeq
In what follows, we refer to the theorem proved in Ref.~\cite{Borde} as the BGV theorem.

Even though the BGV theorem is sometimes called a "singularity theorem", it does not imply the existence of spacetime singularities.  All it says is that an expanding region of spacetime cannot be extended to the past beyond some boundary ${\cal B}$.  All past-directed timelike and null geodesics, except perhaps a set of measure zero, reach this boundary in a finite proper time (finite affine parameter in the null case).  It follows that inflation alone is not sufficient to provide a complete description of the universe, and some new physics is necessary to determine the boundary conditions on ${\cal B}$.  In this sense, it was concluded in \cite{Borde} that inflation must have had some sort of a beginning.\footnote{It was shown in \cite{Borde} that  the same conclusion applies to the `cyclic universe' models \cite{Steinhardt}, which also satisfy the average expansion condition (\ref{H>0}).}
  
An important characteristic of the boundary condition on ${\cal B}$ can be deduced from the second law of thermodynamics.  The entropy of the presently observable part of the universe is many orders of magnitude lower than its maximum value.  
And the second law tells us that the initial entropy of our comoving region on ${\cal B}$ must be lower still.  This suggests that the universe must have originated in a very special (non-random) state of extremely low entropy \cite{Penrose,Wald}.

The validity of the BGV theorem is not in question, but its interpretation has generated some controversy.    
Linde \cite{Linde1} emphasized that the theorem still allows some geodesics (a set of measure zero) to be past-eternal. A simple example is a `comoving' geodesic ${\bf x}={\rm const}$ in de Sitter space with flat spatial slicing,
\beq
ds^2 = dt^2 - e^{2Ht} d{\bf x}^2.
\label{flat}
\eeq
Observers evolving along such geodesics will see inflation continue from the infinite past.  This is true, but all other past-directed geodesics reach $t=-\infty$ in a finite proper time.  The null surface $t=-\infty$ plays the role of the boundary ${\cal B}$ in this example.  We say that inflation must have a beginning in the sense that some physical process has to enforce the boundary conditions on that surface (or on some surface in its future).

Susskind \cite{Susskind1,Susskind2} has pointed out that observers in an eternally inflating universe will predominantly live in the far future of the boundary surface ${\cal B}$.  As a result, the boundary conditions on ${\cal B}$ will be all but forgotten.  The only exception is the `persistence of memory' effect \cite{persistence} -- the asymmetry in the distribution of bubbles colliding with our bubble over the sky.  This asymmetry is related to the orientation of the boundary surface and is potentially observable.  Apart from this effect, observers will not be able to detect any evidence for the beginning.  This point, however, does not contradict the BGV theorem, which makes no claims about the observability of the boundary surface.

A number of authors emphasized that the beginning of inflation does not have to be the beginning of the universe.  The `emergent universe' scenario \cite{Mulryne,Sergio,Yu,Barrow,Graham} assumes that the universe approaches a static or oscillating regime in the asymptotic past.  In this case, the average expansion rate is $H_{av}=0$, so the condition (\ref{H>0}) is violated.  The problem with this scenario is that static or oscillating universes are generally unstable with respect to quantum collapse and therefore could not have survived for an infinite time before the onset of inflation \cite{Dabrowski,Mithani1,Mithani2}.

A completely different way to avoid a beginning of the universe was proposed by Aguirre and Gratton \cite{Aguirre1,Aguirre2,Aguirre3}, who argued that inflation may be eternal both to the past and to the future if the thermodynamic arrow of time is allowed to point in opposite directions in different spacetime regions.  This can be illustrated using de Sitter space with closed spatial foliation as an example,
\beq
ds^2 = d\tau^2 - H^{-2}\cosh^2(H\tau) d\Omega_3^2.
\label{closed}
\eeq
The time coordinate $\tau$ varies monotonically from $-\infty$ to $+\infty$, with the universe contracting at $\tau<0$, bouncing at $\tau=0$, and re-expanding at $\tau>0$.  There is no contradiction with the BGV theorem, since the eternal expansion of the universe is preceded by contraction.

Suppose the vacuum that fills this de Sitter space is a metastable (false) vacuum and that it can decay to one or more lower-energy vacua through bubble nucleation.  Suppose further that we impose a boundary condition that the entire universe is in a false vacuum state in the asymptotic past, $\tau\to-\infty$.  Then bubbles nucleating at $\tau\to-\infty$ will fill the space, the energy in the bubble walls will thermalize, and the universe will contract to a big crunch and will never get to the bounce and to the expanding phase.

Suppose now that the low-entropy boundary condition of false vacuum is imposed at the bounce surface, $\tau=0$.  
More specifically, we can require that the spacetime should include a spacelike surface ${\cal B}$ with the following properties: (i) the entire surface is in the state of a de Sitter false vacuum of energy density $\rho_F$;  (ii) it has the geometry of a 3-sphere of radius $R=(8\pi G\rho_F/3)^{-1/2}$ and zero extrinsic curvature.  Then the metric in the vicinity of ${\cal B}$ has the form of Eq.~(\ref{closed}), with ${\cal B}$ at $\tau=0$.  Aguirre \& Gratton argue that with such boundary conditions, the thermodynamic arrow of time would point away from the bounce surface, that is, to the future at $\tau>0$ and to the past at $\tau<0$ (see Fig.~1a).  Since both the boundary conditions on ${\cal B}$ and the laws of physics that govern time evolution are time-symmetric, the evolution at $\tau < 0$ is statistically (that is, up to the variation in random locations of bubble nucleation centers) a time-reversed version of the evolution at $\tau >0$.  In particular, bubbles formed at $\tau < 0$ are expanding in the negative time direction.

\begin{figure}[h]
\begin{center}
\includegraphics[width=9cm]{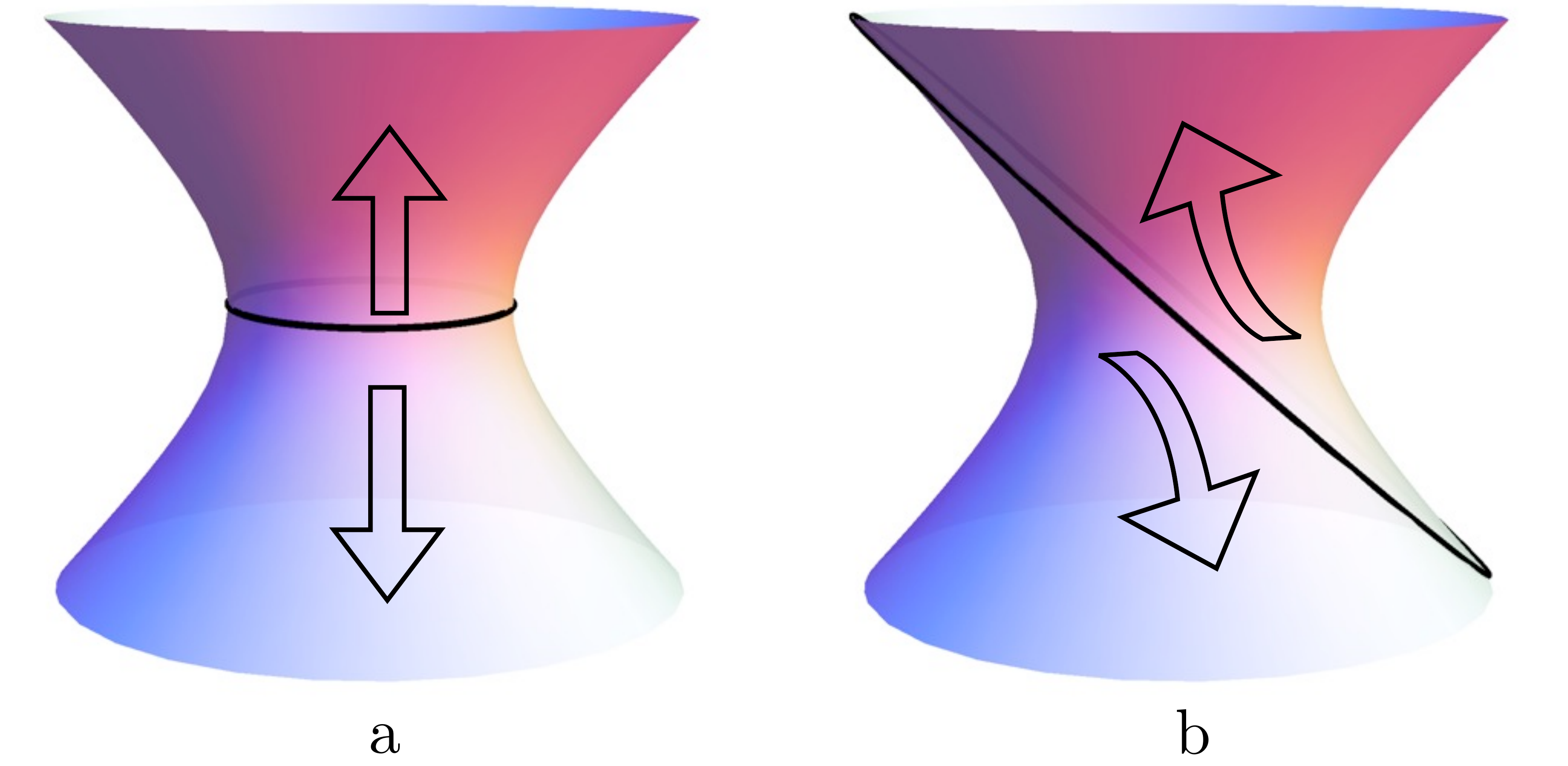}
\caption{Arrows of time in the Aguirre-Gratton scenario with a low-entropy boundary condition (a) at the de Sitter bounce and (b) on the null hypersurface $t=-\infty$.}
\end{center}
\end{figure}

The bounce surface in de Sitter space is defined up to boost transformations.  The surfaces obtained from one another by such transformations are completely equivalent, but in the limit of infinite boost the surface becomes null.  It then coincides with the $t=-\infty$ surface for the appropriately chosen flat foliation (\ref{flat}).  This gives an alternative version of the Aguirre-Gratton (AG) scenario \cite{Aguirre1,Aguirre2,Aguirre3}, with the arrows of time indicated in Fig.~1b.

Even though the spacetime has no boundary in the AG model, it does include a surface ${\cal B}$ on which the low-entropy (vacuum) boundary condition must be enforced by some unknown mechanism.   This Cauchy surface of minimum entropy plays the role of the beginning of the universe in this scenario.  

Instead of a boundary condition at the bounce, a bidirectional arrow of time can be enforced by imposing some global conditions on the spacetime.  Aguirre \cite{Aguirre3} suggested a condition that he called Consistent Cosmic Censorship (CCC): {\it Past singularities cannot be observed by any physical observer.}\footnote{A similar idea has been proposed by Page \cite{Pagenobounce} in the context of quantum cosmology.  He suggested that the universe is in a mixed state comprised of bouncing spacetimes with asymptotic de Sitter regions in both time directions.  Page later rejected this particular proposal on phenomenological grounds \cite{Pagebounce}, but some version of it may still be used to specify the quantum state of the universe with a bidirectional arrow of time.}   This generalizes Penrose's (strong) Cosmic Censorship conjecture \cite{censorship}, requiring (roughly speaking) that observers cannot see past singularities other than the big bang.  Note that future singularities are not excluded: observers can hit black hole singularities or big crunches inside Anti-de Sitter bubbles.  In a universe with a bidirectional arrow of time, black holes and bubbles prior to the bounce are time-reversed versions of black holes and bubbles formed after the bounce.  A post-bounce observer will generally have such pre-bounce singularities in his chronological past, but Aguirre argues that physical observers can observe (that is, gain information about) only events in their thermodynamic past (that is, after the minimum entropy hypersurface).  The class of spacetimes allowed by CCC is considerably wider than that specified by the AG boundary condition.  

A more ambitious version of the AG model has been proposed by Carroll and Chen \cite{CarrollChen,CarrollChen1,CarrollChen2,Carroll}.  They suggest that some `generic' initial conditions are imposed on a spacelike Cauchy surface ${\cal B}$ and argue that the entropy of any comoving region on ${\cal B}$ is less than its maximum value, simply because the comoving entropy is unbounded from above.  Then the second law requires that, apart from thermal fluctuations, the entropy grows both to the past and to the future of ${\cal B}$.  Assuming a stable true vacuum with a positive energy density, Carroll and Chen (CC) argue that the resulting spacetime will be non-singular, apart from the singularities that can be formed in black hole interiors, and that the universe will exhibit eternal inflation in both time directions.  They call this scenario "spontaneous inflation".  An important difference from the AG model is that CC claim that the boundary conditions on ${\cal B}$ do not have to be special: pretty much any boundary conditions would suffice.  Similar ideas have been more recently discussed by Guth \cite{GuthArizona}.\footnote{The idea that inflation would naturally arise from random initial conditions was first introduced by Linde \cite{chaotic}, but he discussed only the evolution in the forward time direction.} 

In the present paper I will examine several aspects of cosmologies with a bidirectional thermodynamic arrow of time.  The assumption that bubbles nucleating prior to the Cauchy hypersurface ${\cal B}$ expand in the negative time direction appears to be somewhat counter-intuitive, and I begin in the next Section with a simple $(1+1)$-dimensional model, which is closely analogous to the AG scenario and for which explicit quantum states can be constructed.  I conclude that bubbles on opposite sides of a de Sitter bounce indeed expand 
in opposite time directions if a vacuum boundary condition is imposed at the bounce.  Such a boundary condition could originate from quantum cosmology and I discuss its different versions, resulting from the tunneling and Hartle-Hawking wave functions of the universe.  In Section III I turn to the CC scenario and show, with relatively mild assumptions, that the Cauchy surface in a non-singular (apart from black holes) universe with asymptotically inflating regions must necessarily be compact.  In other words, such a universe must be closed.  I also argue that the size of the universe at the bounce between the two asymptotic inflating regions cannot much exceed the de Sitter horizon.  The resulting spacetime structure is thus very 
similar to that in the AG scenario; it does require low-entropy boundary conditions at the bounce.  The conclusions of the paper are summarized and discussed in Section IV; in particular, I discuss how the conclusions are modified by allowing cosmological singularities.  In this case, an infinite Cauchy surface with `random' initial data will generally produce inflating regions in both time directions.  These regions, however, will be surrounded by singularities and will have singularities in their past or future.

\section{de Sitter bounce scenario}

\subsection{Pair production analogy}

To get a better intuitive grasp of the AG scenario, it may be useful to consider a simple model of pair production in a $(1+1)$-dimensional de Sitter space with a constant electric field.  The metric can be written as
\beq
ds^2 = dt^2 - H^{-2}\cosh^2(Ht) d\chi^2,
\label{2d}
\eeq
with $0\leq \chi < 2\pi$, and the electric field is
\beq
F^{\mu\nu}=\epsilon^{\mu\nu}\frac{E}{\sqrt{-g}}.
\label{F}
\eeq
Here, $\epsilon^{\mu\nu}$ is an antisymmetric tensor with $\epsilon^{01}=1$, and $g$ is the determinant of the metric tensor.  The electric field in $(1+1)D$ is similar to a cosmological constant.  Maxwell equations require that $E={\rm const}$, so
\beq
F^{\mu\nu}F_{\mu\nu} = -2E^2 = {\rm const},
\eeq
and the energy-momentum tensor has a vacuum form, $T_{\mu\nu}\propto g_{\mu\nu}$.

We shall assume that the electric field is coupled to a quantum field $\psi$ of charge $e$ and mass $m$.  Then particle-antiparticle pairs will be produced by the Schwinger process.  Pair production in $(1+1)$ dimensions is closely analogous to bubble nucleation, with particles of the pair playing the role of bubble walls.  (In flat spacetime, this analogy has been recently explored in Refs.~\cite{Sugumi1,Sugumi2}.)  At $\tau> 0$, once a pair is created, the positively and negatively charged particles are driven by the electric field in opposite directions, resulting in a nonzero expectation value of the electric current $j^\mu$.  The pairs accumulate with time, but they are also diluted by the expansion of the universe.  With a constant expansion rate $H$, one can expect that the current will approach a constant value in the asymptotic future.  (We disregard the back-reaction of the pairs on the electric field.) 

The time evolution of the current is described by the equation
\beq
\nabla_\mu {\tilde j}^\mu = 2e\Gamma,
\label{jC}
\eeq
where
\beq
{\tilde j}^\mu = \frac{\epsilon^{\mu\nu}j_\nu}{\sqrt{-g}}
\eeq
and $\Gamma$ is the constant pair creation rate per unit length.  The factor 2 is due to the fact that there are two particles in a pair.  Eq.~(\ref{jC}) can be rewritten as
\beq
\epsilon^{\mu\nu}\partial_\mu j_\nu = 2e\Gamma\sqrt{-g},
\eeq
or, assuming that the current is homogeneous, $\partial_1 j^\mu =0$,
\beq
\partial_0 j_1 = 2e\Gamma H^{-1} \cosh (H\tau).
\label{d0j1}
\eeq
The time component of the current is $j_0=0$, since positive and negative charges are created in equal numbers.

The boundary condition requiring that no pairs are present at $\tau=0$ implies
\beq
j_1(\tau=0)=0.
\label{j1=0}
\eeq
The solution of Eq.~(\ref{d0j1}) with this boundary condition is
\beq
j_1 = 2e\Gamma H^{-2}\sinh (H\tau).
\label{j1}
\eeq
The invariant magnitude of the current can be defined as
\beq
J \equiv \sqrt{-j_\mu j^\mu} = 2e\Gamma H^{-1} \tanh (H\tau).
\label{J}
\eeq
The current grows from zero at $\tau=0$ and approaches an asymptotic constant value $J=\Gamma/H$ at $\tau\to\pm\infty$.

The concept of a particle is not sharply defined in a time-varying background (see, e.g., \cite{Birrell}), but the ambiguity disappears in the semiclassical limit of large particle mass.  In our case this corresponds to
\beq
m\gg H.
\label{m>>H}
\eeq

Pair production in a constant electric field in de Sitter space has been studied by Garriga \cite{Garriga}.  He used both instanton and Bogoliubov transformation techniques and found complete agreement between the two methods in the semiclassical regime.  In the large mass limit (\ref{m>>H}), the pair production rate is
\beq
\Gamma\sim \frac{Q}{2\pi} \exp\left[-\frac{2\pi}{H^2}(Q-eE)\right],
\label{Gamma}
\eeq
where
\beq
Q=(m^2H^2 +e^2E^2)^{1/2}.
\label{B}
\eeq
For $eE\gg mH$, particle separation in the pairs, $d \sim 2m/eE$, is much smaller than the Hubble distance $H^{-1}$, and Eq.~(\ref{Gamma}) reduces to the Schwinger pair production rate in flat space \cite{Schwinger}.  

Garriga \cite{Garriga} found the quantum state explicitly, by finding the mode functions for a charged massive scalar field in de Sitter space in the presence of an electric field.  He used the coordinates (\ref{flat}), which cover only half of de Sitter space, so his analysis cannot be directly applied to our situation.  It may also be possible to construct an explicit time-symmetric quantum state in the $(1+1)$-dimensional version of the closed de Sitter metric (\ref{closed}).  The choice of the mode functions in \cite{Garriga} corresponds to the in-vacuum at $t\to -\infty$.  This appears to be analogous to imposing the AG boundary condition on the null hypersurface $t=-\infty$.  It would be interesting to check if a non-singular quantum state can be constructed by supplementing Garriga's quantum state in the upper half of de Sitter space with a time-reversed state in the lower half.  

Another interesting case is that of massless spin-1/2 particles.  The $(1+1)D$ universe is then similar to a superconducting cosmic string with fermionic charge carriers \cite{Witten}, and the pair creation rate in Eq.~(\ref{jC}) can be found exactly,
\beq
\Gamma=\frac{eE}{2\pi}. 
\eeq
In fact, ${\tilde j}^\mu$ in this case plays the role of axial current, and Eq.~(\ref{jC}) is a statement of axial anomaly \cite{Peskin}. 

The solution (\ref{j1}) for the current has the property
\beq
j_1(-\tau)=-j_1(\tau),
\eeq
indicating that the number of pairs is unboundedly large in the asymptotic past.  These pairs annihilate during the contracting phase of the universe, resulting in a state with no pairs at $\tau = 0$.  This would require an extreme fine-tuning if the boundary conditions were set at $\tau\to -\infty$.  However, the state with current growing in both time directions is naturally obtained if we set the boundary condition (\ref{j1=0}) at $\tau=0$.  The evolution backwards in time from $\tau=0$ is then statistically equivalent to the evolution forward in time.  We expect that a similar global structure would arise in a $(3+1)D$ universe with the AG boundary conditions at the bounce.

We note finally that the boundary condition (\ref{j1=0}) breaks the de Sitter symmetry of the spacetime.  In the absence of pair production, all spatial slices obtained by boosting the slice at $\tau = 0$ are equivalent.  But if the vacuum is metastable, the symmetry is spontaneously broken by pair production.

\subsection{Boundary conditions from quantum cosmology}

A low-entropy boundary condition at the bounce may naturally arise from quantum cosmology, where the entire universe is treated quantum-mechanically and is described by a wave function $\psi$.  This wave function is defined on superspace, which is the space of all 3-geometries and matter field configurations, and satisfies the Wheeler-DeWitt equation
\beq
{\cal H}\psi = 0,
\eeq
where ${\cal H}$ is the Hamiltonian density operator.  The form of the boundary conditions for $\psi$ (which are not to be confused with the spacetime boundary conditions on ${\cal B}$) is a subject of continuing debate.  The two best studied proposals are the tunneling \cite{LindeRep,Zeldovich,Rubakov,AV84,tunneling} and the Hartle-Hawking \cite{HH} proposals; they lead to rather different pictures for the origin of the universe. 

In the  semiclassical regime, the wave function can be represented as a superposition of WKB-type terms, 
\beq
A e^{iW}.
\label{WKB}
\eeq
Here, $W$ is the Hamilton-Jacobi function, describing a congruence of classical trajectories (histories).  A time-reversed congruence is described by $A^* e^{-iW}$.  The tunneling approach requires that $\psi$ should include only outgoing waves at the boundaries of superspace.  For a semiclassical wave function, this means that only terms with trajectories evolving towards the boundary should be included \cite{tunneling}.  In particular, only expanding universes are allowed in the limit of large volume.  For a bouncing de Sitter-type universe, the contracting phase is absent, so the universe nucleates spontaneously at the bounce point.

The instanton describing this nucleation process is the Euclideanized de Sitter space (a 4-sphere).  Its Euclidean action is $S_E=-3/8G^2\rho_v$, where $\rho_v$ is the vacuum energy density, and the corresponding amplitudes in the tunneling wave function are suppressed by the factor 
\beq
\exp\left(-|S_E|\right)  = \exp\left(-\frac{3}{8G^2\rho_v}\right).
\label{tunneling}
\eeq
The dominant contribution to this wave function is given by histories originating in the highest-energy vacuum of the underlying particle theory.\footnote{If the spectrum of possible values of $\rho_v$ extends all the way to the Planck scale, $\rho_{Pl}=G^{-2}$, the suppression factor (\ref{tunneling}) becomes ${\cal O}(1)$, and the universe can originate with a comparable probability in any of the vacua with $\rho_v\sim\rho_{Pl}$.}  Furthermore, midi-superspace analysis with the inclusion of linearized perturbations indicates that all quantum fields at the bounce are in de Sitter-invariant (Bunch-Davies) vacuum states  \cite{Vachaspati}.  This specifies a very low-entropy state at the beginning of the universe.  

In the Hartle-Hawking approach, the wave function is real, and for each WKB term (\ref{WKB}) there is a complex conjugate term of the same magnitude.  The exponential suppression factor (\ref{tunneling}) is replaced in this case by
\beq
\exp\left(-S_E\right)  = \exp\left(+\frac{3}{8G^2\rho_v}\right),
\label{HH}
\eeq
so the wave function is dominated by bouncing histories characterized by the {\it lowest} positive vacuum energy density.
Once again, perturbative analysis indicates that the quantum state at the bounce is the de Sitter-invariant vacuum state \cite{Halliwell}.  Apart from the sign in the exponent, the key difference from the tunneling case is that both expanding and contracting histories are now represented in $\psi$ with equal amplitudes. The simplest interpretation of this is that the Hartle-Hawking wave function describes a contracting and re-expanding universe with the AG boundary conditions at the bounce \cite{HertogHartle}.\footnote{This kind of boundary condition has been discussed in Ref.~\cite{Hartle} in the context of generalized quantum mechanics.}  

We conclude that quantum cosmology can provide low-entropy (vacuum) boundary conditions, both for a universe spontaneously nucleating from "nothing" (with the tunneling wave function) and for a contracting and re-expanding universe with a bidirectional arrow of time (with the Hartle-Hawking wave function).  In the former case, the universe is most likely to begin in the highest-energy vacuum, while in the latter case the bounce is most likely to occur in the lowest (positive) energy vacuum.  

Before we move on to the next subject, I would like to comment on some other relevant proposals for the wave function of the universe.  The proposal of Ref.~\cite{Linde84} gives a wave function including both expanding and contracting branches in a FRW minisuperspace model.  However, the behavior of this wave function in the classically forbidden region does not correspond to a contracting and bouncing universe: instead of decaying, it grows away from the turning point \cite{discord}.  The physical interpretation of this wave function is therefore unclear.

Page \cite{Pagebounce} proposed that the wave function of the universe is given by a superposition of semiclassical states describing bouncing cosmologies.  The corresponding spacetimes are assumed to be approximately de Sitter near the bounce, with all quantum fields in approximately Bunch-Davies states and with the thermodynamic arrow of time pointing away from the bounce in both time directions.  The main difference from the Hartle-Hawking wave function is that the amplitude for the semiclassical variables (the radius of the universe and the homogeneous component of the inflaton field) is given by a special prescription (using DeWitt's minisuperspace metric), rather than by the instanton action, as in Eq.~(\ref{HH}).  By constructtion, each semiclassical component of the Page's wave function describes a bouncing universe with the AG boundary conditions at the bounce.  The problem with this prescription is that it is rather {\it ad hoc} and is defined only approximately.  It is not clear how it can be made precise or extended to a more general class of models.

\section{Spontaneous inflation}

\subsection{Topology of the universe}

We now turn to the spontaneous inflation scenario of Carroll and Chen (CC).   Its starting point is that some `generic', regular boundary conditions are specified on a spacelike Cauchy surface ${\cal B}$.  This initial state is then evolved both to the past and to the future of ${\cal B}$.  Assuming that the lowest-energy (true) vacuum has a positive energy density $\rho_T >0$, CC argue that the universe approaches an empty true-vacuum de Sitter configuration in both time directions.  Islands of inflating high-energy vacua can then form spontaneously by quantum fluctuations.  This can occur either through false-vacuum bubble nucleation \cite{LeeWeinberg,recycling} or through quantum creation of baby universes by Farhi-Guth-Guven-type tunneling \cite{FarhiGuth}.\footnote{A true vacuum with $\rho_T >0$ may be problematic due to the Boltzmann brain menace.  It has been argued in \cite{Dyson,Albrecht,Page} that in such a vacuum a region like our observable part of the universe is much more likely to arise as a `thermal' de Sitter fluctuation than as a result of a large quantum jump to a high-energy vacuum with subsequent inflation.  Disembodied brains, deluded to believe that they are observing a region like ours, will occur with a still higher probability. In the context of eternal inflation, the numbers of Boltzmann brains and of normal observers are both infinite, and their relative abundance depends on how infinities are regulated, that is, on the choice of measure.  The measure problem still remains unresolved (see, e.g., \cite{Freivogel} for a review), but the phenomenologically favored scale factor and lightcone measures both predict Boltzmann brain dominance in the case of $\rho_T >0$ \cite{Salem,Bousso}.  This difficulty, however, may not be very severe.  As Carroll and Chen point out, the stability condition for the low-energy vacuum can be relaxed, allowing it to decay by nucleating anti-de Sitter or Minkowski bubbles.  They argue that the scenario remains essentially unchanged if the bubble nucleation rate is sufficiently low.}

The claim that a generic state on the Cauchy surface ${\cal B}$ evolves to the true vacuum is far from being obvious.  
For example, contracting regions on ${\cal B}$ may collapse to a singularity.  To avoid this danger, CC introduce an additional assumption -- that the surface ${\cal B}$ is infinite.  They argue that in this case any contracting region will be embedded in a larger expanding region, so the contracting regions will form isolated black holes, which will eventually evaporate.  (Note that the black holes in the past of ${\cal B}$ are time-reversed versions of the usual black holes.)  The resulting spacetime structure is illustrated in Fig.~2.  This argument is not entirely convincing: if contracting regions are distributed at random and occupy more than about 30\% of the volume on ${\cal B}$, they will percolate, forming an infinite collapsing region, rather than isolated black holes.  

There is also a more serious problem.   We will show that, under rather mild assumptions, a spacetime with an infinite Cauchy surface envisioned in the CC scenario does not exist.  The proof relies on the Penrose-Hawking global techniques (for a review see \cite{HawkingEllis,WaldBook}).  The assumptions are: (i) the existence of a global Cauchy surface, (ii) past null completeness, and (iii) the null convergence condition. 

\begin{figure}[h]
\begin{center}
\includegraphics[width=11cm]{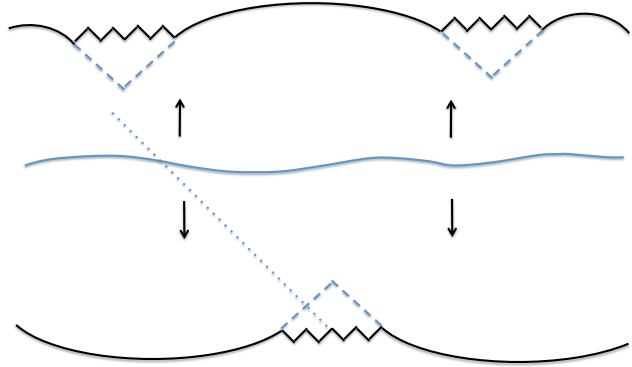}
\caption{Spacetime diagram representing the Carroll-Chen scenario.  The Cauchy surface is shown by a blue curved line.   Black hole (and time-reversed black hole) singularities, shown by zigzag lines, are hidden behind horizons (dashed lines).  The arcs represent spacelike future and past infinities of the inflating regions.  High-energy inflating regions are not shown.  Thermodynamic arrows of time are indicated by arrows.  A past-directed null geodesic terminating at a time-reversed black hole singularity is shown by a dotted line.}
%\label{oscillatingsolution}
\end{center}
\end{figure}

A global Cauchy surface is a spacelike hypersurface such that the entire future and past time development can be determined from the data on that surface.  The existence of a global Cauchy surface is already assumed in the CC scenario, so this is not an independent assumption.  Past null completeness means that all null geodesics can be extended to the past without limit.  This condition is often used as a standby for the absence of spacetime singularities.  We shall indicate below how it can be relaxed to allow for the possibility of black hole formation in the CC scenario.  The null convergence condition (NCC) requires that
\beq
R_{\mu\nu}N^\mu N^\nu \geq 0
\label{NCC}
\eeq
for all null vectors $N^\mu$, where $R_{\mu\nu}$ is the Ricci tensor.  Combined with Einstein's equations, NCC is equivalent to the null energy condition (NEC), requiring that
\beq
T_{\mu\nu}N^\mu N^\nu \geq 0
\eeq
for all null $N^\mu$.  
NEC is satisfied by all known forms of classical matter, but it can be violated by quantum effects (see, e.g. Ref.~\cite{Ford} and references therein).  In particular, it is only marginally satisfied for the vacuum form of the energy-momentum tensor, $T_{\mu\nu}=\rho_v g_{\mu\nu}$, and can be violated by quantum fluctuations during inflation \cite{BordeV,Vachaspati2}.  On the other hand, quantum effects of this kind are only essential in localized spacetime regions, where high-energy inflating islands are nucleated. If the key premise of the CC scenario -- that a generic initial state evolves to a true vacuum in both time directions -- is correct, it should hold even in the absence of high-energy islands, so its validity can be tested in models where such islands do not form.  For example, models where high-energy inflation is not possible should still yield a spacetime structure shown in Fig.~2.  In what follows, we shall disregard possible quantum violations of the NCC.\footnote{Note also that the theorem based on NCC that we are going to use below can also be derived using weaker conditions, such as the integral convergence condition \cite{Tipler}, repeated integral convergence condition \cite{Borde2}, or the generalized second law \cite{Wall}.  These integral conditions are consistent with some localized violations of the NCC, although it remains to be checked that they apply in our context here.} 

The starting point of the argument\footnote{Here I follow the discussion in Ref.~\cite{Wall}, where the compactness of the Cauchy surface is proved using the generalized second law.} is the observation that the asymptotic de Sitter region to the future of ${\cal B}$ necessarily contains past-trapped surfaces.  (A past-trapped surface ${\cal S}$ is a $2D$ surface such that both outward and inward directed systems of null geodesics emanating orthogonally from ${\cal S}$ toward the past are converging at all points on ${\cal S}$.)  For example, any spherical surface lying on a $t={\rm const}$ spatial slice in coordinates (\ref{flat}) and having radius bigger than the de Sitter horizon $H^{-1}$ is past-trapped.  With the assumptions (i)-(iii), it can then be shown that the Cauchy surface ${\cal B}$ must be compact.  This is a standard result (see, e.g., Refs.~\cite{HawkingEllis,Borde1,Wall}), so only an outline of the proof will be given here.

Suppose ${\cal S}$ is a past-trapped surface.  The past light cone of ${\cal S}$, $E^-({\cal S})$, can be defined as the boundary of its past, $E^-(S)={\dot I^-}(S)$. In the vicinity of ${\cal S}$, the light cone is comprised of the two past-directed sheets of null geodesics emanating from ${\cal S}$.  The null geodesics on $E^-({\cal S})$ converge, and it follows from NCC and past null completeness that each of these geodesics comes to a conjugate point (that is, crosses nearby geodesics on $E^-({\cal S})$) in a finite affine parameter time.  After crossing, the geodesics do not stay on the light cone and enter its interior.  Since this happens to all geodesics in a finite affine time, the light cone must be compact.  And since $E^-({\cal S})$ is a boundary of a set $(I^-({\cal S}))$, it must have no boundary.  It must also be achronal, which means that no two points on $E^-({\cal S})$ can be connected by a timelike curve.  (Otherwise, one of these points would be in the interior of $I^-({\cal S})$.)  

Now, the existence of a compact, edgeless, achronal hypersurface is inconsistent with a non-compact Cauchy surface.  
In order to see this, consider a smooth timelike vector field $V^\mu(x)$ whose integral curves cross the Cauchy surface ${\cal B}$ exactly once.  (The existence of such a field follows from the fact that ${\cal B}$ is a Cauchy surface.)  Since the light cone $E^-({\cal S})$ is achronal, the integral curves of $V^\mu$ can cross it no more than once.  Thus they define a continuous one-to-one map $E^-({\cal S})\to {\cal B}$.  Such a map, however, is possible only if ${\cal B}$ is itself compact.

The above proof can be extended to allow for black holes \cite{VilWall}, which are expected to form in the CC scenario, as the initial data on ${\cal B}$ is evolved in both time directions.  In this case, we cannot assume that all past-directed null geodesics can be continued indefinitely, since some of them will run into (time-reversed) black hole singularities.  (A past-directed null geodesic terminating at a singularity is shown by a dotted line in Fig.~2.)

We shall assume, however, in keeping with the spirit of cosmic censorship, that all singularities are enclosed inside of isolated black hole horizons.  Then
we can join the light cone $E^-({\cal S})$ with the horizon surfaces at their intersections, resulting in a new compact, edgeless, achronal hypersurface.\footnote{Here I assume that the past-trapped surface ${\cal S}$ is not completely contained inside one of the (time-reversed) black hole horizons.  Otherwise, this configuration would represent an inflating universe contained inside of a black hole.  Such objects can exist in an asymptotically flat background, as discussed in \cite{Blau} and references therein.  But in the presence of a positive cosmological constant, the black hole exterior region should be asymptotically de Sitter and should also contain past-trapped surfaces.  Then, following the same argument as above, we conclude once again that the Cauchy surface must be compact.}
This allows us to conclude, as before, that the Cauchy surface ${\cal B}$ must be compact.  
The details of this argument will be presented elsewhere \cite{VilWall}.

The conclusion is that the universe must be closed in the CC scenario -- which means that the spacetime must contain a closed spacelike hypersurface.  For such a universe, it seems rather unlikely that it will expand forever and approach an empty true vacuum state in both time directions, starting from generic initial conditions.  We know that a closed FLRW universe filled with ordinary matter recollapses to a big crunch.  
The crunch can be avoided in the presence of a positive vacuum energy density $\rho_T$ if the universe is expanding fast enough, so that matter density is diluted below $2\rho_T$
before the onset of collapse.  But then there is a danger that the universe will collapse to a big crunch in the opposite time direction.   Alternatively, we could start the universe with a slow expansion rate and arrange the matter density to be lower than $2\rho_T$, so that gravity of matter is dominated by the repulsive gravity of the vacuum from the start.  But in this case   
the resulting spacetime resembles the AG picture of Fig.~1(a), with a low-entropy boundary condition at the bounce.  I will make these considerations more quantitative in the next subsection.  I will also argue that the size of the universe at the bounce should not much exceed the de Sitter horizon.

Deviations from FLRW geometry appear to make things only worse.  Suppose we introduce a small density perturbation in the future asymptotic de Sitter region.  As we follow the evolution to the past, the perturbation will grow and will become ${\cal O}(1)$ well before we get to the bounce region, so the universe will collapse to a big crunch. The same argument, of course, applies to perturbations in the past asymptotic de Sitter region.  It is therefore highly unlikely for these regions to be connected by a smooth bounce, unless we impose very special boundary conditions in the bounce region.  Replacing small perturbations by large deviations from homogeneity and isotropy would hardly make this more probable.

\subsection{Size at the bounce}

To assess the size of the universe at the bounce, let us first consider a FLRW version of the model,
\beq
ds^2 = d\tau^2 -a^2(\tau) d\Omega_3^2.
\eeq
The scale factor $a(\tau)$ satisfies the equation
\beq
{\dot a}^2 + 1 = \frac{8\pi G}{3}\rho a^2,
\label{Friedmann1}
\eeq
where $\rho = \rho_T + \rho_m$, $\rho_T >0$ is the true vacuum energy density, and $\rho_m \geq 0$ is the density of matter.
Since $a(\tau)$ grows without bound at $\tau\to\pm\infty$, it must have a minimum at some finite time $\tau_*$.  At that time, ${\dot a}(a_*)=0$ and Eq.~(\ref{Friedmann1}) gives
\beq
a_{min}\equiv a(\tau_*) =\left(\frac{3}{8\pi G\rho(\tau_*)}\right)^{1/2} \leq H_T^{-1}, 
\label{amin}
\eeq
where
\beq
H_T = \left(\frac{8\pi G\rho_T}{3}\right)^{1/2}
\eeq
is the expansion rate of the true vacuum.

An additional constraint on $a_{min}$ can be obtained from the condition ${\ddot a}(\tau_*)>0$.  Assuming that the matter pressure is non-negative, $P_m\geq 0$, it is easy to show that $a_{min}\geq 3^{-1/2} H_T^{-1}$.
Thus, the scale factor at the bounce must be close to the true vacuum de Sitter horizon $H_T^{-1}$.

Without assuming FLRW symmetry, a bounce can be defined as a spacelike hypersurface $\Sigma$, such that ${\dot V} = 0$ and ${\ddot V}  >0$, where $V$ is the volume of $\Sigma$ and overdots stand for derivatives with respect to proper time along the geodesic congruence orthogonal to $\Sigma$.  For example, ${\dot V}$ is the derivative of the volume $V$ as each point of $\Sigma$ is moved the same distance along the congruence.  I was not able to derive any bound on the size of the bounce in this most general case.   Below I give some partial results in that direction.

Let us first consider a time-symmetric bounce, which can be defined as a spacelike hypersurface ${\cal B}$ of vanishing extrinsic curvature.   Then the intrinsic scalar curvature of ${\cal B}$ can be found from the Hamiltonian constraint,
\beq
^{(3)}R=16\pi G \rho > 16\pi G \rho_T,
\label{3R}
\eeq
where $\rho = T_{\mu\nu}u^\mu u^\nu$ and $u^\mu$ is a unit vector normal to ${\cal B}$.  If ${\cal B}$ is a 3-sphere of radius $a_{min}$, then $^{(3)}R = 6 a_{min}^{-2}$, and (\ref{3R}) reduces to Eq.~(\ref{amin}).

A bound on the scalar curvature (\ref{3R}) is not generally sufficient to place an upper bound on the size of the corresponding spatial section.  Such a bound can be derived for 2-dimensional surfaces using the Gauss-Bonnet theorem,
\beq
\int dA ~{^{(2)}R} = 4\pi \chi,
\label{GaussBonnet}
\eeq
where $dA$ is the area element, $\chi = 2 - 2g$ is the Euler characteristic, and $g$ is the genus of the surface.  For $^{(2)}R> {^{(2)}R}_{min}>0$, the surface must have the topology of a sphere $(g=0)$, and its total area is bounded by
\beq
A < 8\pi / {^{(2)}R}_{min}.
\eeq

In 3 dimensions, a bound on the size of the bounce section ${\cal B}$ can be obtained if the Ricci curvature is bounded below,
\beq
^{(3)}R_{ij} n^i n^j \geq C > 0,
\label{Riccibound}
\eeq
for an arbitrary unit 3-vector $n^i$ in ${\cal B}$. The quantity $^{(3)}R_{ij} n^i n^j$ has the meaning of the average sectional curvature, where the averaging is over $2D$ sections passing through the vector $n^i$.  Myers theorem \cite{Myers} states that if (\ref{Riccibound}) is satisfied, then the geodesic distance $d$ between any two points in ${\cal B}$ does not exceed $\pi \sqrt{2/C}$.  Assuming that $C\sim {^{(3)}R}_{min} \sim G \rho_T$, this gives
\beq
d\lesssim H_T^{-1}.
\eeq

As of now, a bounce at a very large volume, $V\gg H_T^{-3}$, cannot in principle be excluded, if large deviations from homogeneity and isotropy are allowed. This possibility, however, appears rather remote: a highly inhomogeneous or anisotropic universe is likely to exhibit singular behavior.  For example, the Kantowski-Sachs metric, 
\beq
ds^2 = dt^2 - a^2(t) dz^2  - b^2(t) d\Omega_2^2 ,
\eeq
can be thought of as an ellipsoidal universe with two equal axes, in the limit when the third axis is much larger than the other two.
It has been shown in \cite{SolomonsDunsbyEllis,Pagenobounce} that a bounce with ${\ddot b}>0$ is impossible in this model, as long as the weak energy condition is satisfied.

\section{Summary and discussion}

In this paper, we discussed cosmological scenarios with a bidirectional arrow of time.  The Aguirre-Gratton (AG) scenario assumes a de Sitter-like bounce with the thermodynamic arrow of time pointing in opposite directions away from the bounce.
I argued that such a scenario may naturally arise in quantum cosmology with the Hartle-Hawking wave function of the universe.  This choice of the wave function favors a de Sitter-like bounce in a vacuum state of the lowest positive energy density.  The tunneling wave function, on the other hand, suggests that semiclassical spacetime is present only in one time direction from the bounce and favors the initial vacuum of the highest energy density.

Even though the spacetime has no boundary in the AG model, it does include a hypersurface on which the low-entropy (vacuum) boundary condition must be enforced by some mechanism.   This surface of minimum entropy plays the role of the beginning of the universe in this scenario.  

We also discussed the proposal of Carroll and Chen, that `generic' boundary conditions on a Cauchy surface, with a stable low-energy de Sitter vacuum, naturally yield a bidirectional time development.  They conjecture that, apart from singularities that could develop in black hole interiors, the resulting spacetime is non-singular and has eternally inflating asymptotic regions in both time directions.  Here, I argued that the Cauchy surface in a non-singular (apart from black holes) universe with asymptotically inflating regions must necessarily be compact.  In other words, the universe must be closed, with a bounce at some finite size between the two asymptotic inflating regions.  I also argued that the size of the universe at the bounce cannot much exceed the horizon, so the overall structure of spacetime must be very similar to that in the AG scenario.  The bounce in this kind of spacetime is highly unstable: small perturbations in either asymptotic region destroy the bounce, replacing it with a singularity.  This scenario therefore requires very special boundary conditions at the bounce.      

\begin{figure}[h]
\begin{center}
\includegraphics[width=11cm]{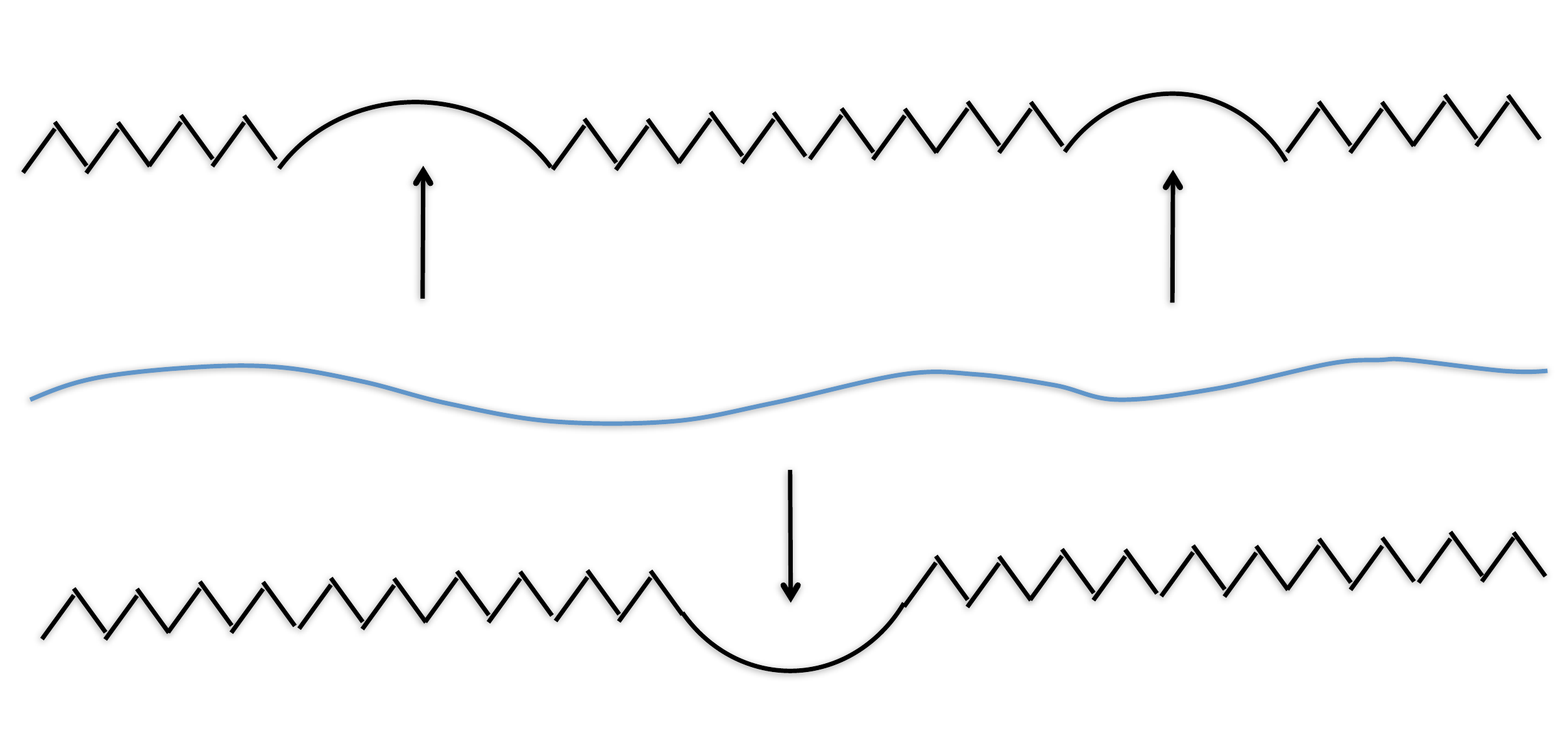}
\caption{Spacetime diagram of a singular universe with an infinite Cauchy surface.  The Cauchy surface is shown by a blue curved line and the singularities by zigzag lines.  The arcs represent spacelike future and past infinities of the inflating regions.  Thermodynamic arrows of time are indicated by arrows.}
%\label{oscillatingsolution}
\end{center}
\end{figure}

This conclusion relies, in an essential way, on the assumption that the spacetime is non-singular, apart from possible black hole singularities.  If this assumption is lifted, then there is no obstruction to having an infinite Cauchy surface.  With generic initial data on this surface, there will be some regions that will inflate towards the future and some regions that will inflate towards the past.  However, regions inflating toward the future will have singularities in their past and vice versa.  There will also be regions with singularities in both time directions; see Fig.~3.  The inflating regions will become sites of eternal inflation, with the thermodynamic arrow of time pointing away from the Cauchy surface.  

The conditions necessary to produce an inflating region are somewhat special (low-entropy).  In the absence of any theory for the initial conditions on the Cauchy surface, this may be difficult to quantify, but it appears that the universe has to be more or less homogeneous and expanding sufficiently fast on a scale considerably greater than the horizon \cite{Piran,Brandenberger,Trodden}. Such regions will be relatively rare on the initial value surface, so we expect inflating regions to be surrounded by singularities on all sides.  

The spacetime structure in Fig.~3 is rather different from that originally envisioned by Carroll and Chen (Fig.~2), but it does achieve one of the main goals of their scenario -- to produce an inflating universe with an arrow of time from time-symmetric initial conditions.  It is not clear, however, that the conditions assumed on the Cauchy surface ${\cal B}$ can be regarded as generic.  A generic spacelike hypersurface in this kind of spacetime will itself run into singularities, so an infinite regular Cauchy surface appears to be rather special.  Note, by the way, that if one is willing to accept a spacetime besieged by singularities, then the assumption of an infinite Cauchy surface does not seem to be essential.  A large compact Cauchy surface with generic initial data will also yield some inflating regions surrounded by singularities.

It should also be noted that our conclusions rely on the null convergence condition (NCC).  NCC is known to be violated by quantum fluctuations, but such fluctuations do not appear to be essential for our discussion here.  More importantly, violations of NCC may occur in the high curvature regime near classical singularities and may in fact lead to resolution of the singularities (see, e.g., \cite{Ashtekar}).  This may significantly modify the global structure of spacetime \cite{watchers} and may open new possibilities for Carroll-Chen-type scenarios.

Personally, however, I am skeptical about the concept of `random' (or `generic') initial conditions.  I don't think it is a good substitute for a theory of initial conditions, as might for example be given by quantum cosmology.  This concept also appears to be rather ill-defined (as Carroll and Chen acknowledge in their paper).  If indeed the entropy of the universe is unbounded from above, then there is no such thing as a generic (or random, or typical) state.

\subsection*{Acknowledgements}

I am grateful to Anthony Aguirre, Sean Carroll, Alan Guth, Don Page, and Aron Wall for very useful discussions, and to Ben Shlaer for his help with the figures. This work was supported in part by the National Science Foundation (grant PHY-0855447) and by the Templeton Foundation.

\end{document}